\documentclass[12pt]{article}

\begin{document}
\vspace*{-.6in}
\thispagestyle{empty}
\begin{flushright}
CALT-68-2126\\
hep-th/9707119
\end{flushright}
\baselineskip = 20pt

\vspace{.5in}
{\Large
\begin{center}
Remarks on the M5-Brane\footnote{Talk presented at the
{\it Strings '97}}
\end{center}}

\begin{center}
John H. Schwarz\\
\emph{California Institute of Technology, Pasadena, CA  91125, USA}
\end{center}
\vspace{0.1in}

\begin{center}
\textbf{Abstract}
\end{center}
\begin{quotation}
\noindent  The fivebrane of M theory -- the M5-brane -- is an especially interesting
object. It plays a central role in a geometric understanding of the
Seiberg--Witten solution of N=2 D=4 gauge theories as well as in
certain new 6d quantum theories. The low energy effective action is an 
interacting theory of a (2,0) tensor
multiplet. The fact that this multiplet contains a two-form gauge
field with a self-dual field strength poses special challenges. Recent
progress in addressing those challenges is reviewed.\end{quotation}
\bigskip

\section{Super P-Branes}

Type II superstring theories in 10d and M theory in 11d have 32 supercharges.
They admit various BPS $p$-brane configurations that break half of the supersymmetry.
The effective world-volume theory of any of these $p$-branes contains 16 Goldstone
fermion fields, corresponding to the broken supersymmetries, which describe 8
propagating fermionic modes in the world volume. Also, associated to the
broken translation symmetries, there are Goldstone boson fields describing
$D-p-1$ propagating bosonic modes. Since the world-volume theory has
unbroken supersymmetries, it must contain an equal number of bosonic and
fermionic degrees of freedom. 

The discrepancy in the numbers given above is
made up by the addition of $9+p-D$ bosonic modes that correspond to world-volume 
gauge fields. There are three basic cases of interest. The first case is when
no additional degress of feeedom are required, in other words when $D=p+9$.
This is the case for Type IIA or Type IIB strings in 10d or the M2-brane
(also called the supermembrane) of M theory in 11d. In each case there are
eight dimensions transverse to the brane. In the Type IIB case there is actually
an infinite family of strings, labelled by a pair of integers (corresponding
to two-form charges).  Stability requires that these integers are relatively
prime~\cite{jhsa,wittena}. These $(m,n)$ strings transform into one another under the
nonperturbative $SL(2,Z)$ duality group of the IIB theory. 
Building on earlier work that studied the relation between the (1,0)
fundamental string and the (0,1) D string~\cite{tseytlin1,aganagic1},  Townsend has
recently shown~\cite{townsend1} how to construct  world-volume theories for this family
of strings that makes their $SL(2,Z)$ duality properties manifest. The
key step is to introduce an $SL(2,Z)$ doublet of $U(1)$ gauge fields. This is possible,
because a gauge field in 2d is nonpropagating and does not affect the
counting of physical degrees of freedom. 

The second case is when the additional $9+p-D$ bosonic modes are provided by
a $U(1)$ gauge field. Since such a gauge field introduces $p-1$ propagating
degrees of freedom, the counting works precisely when $D=10$, for all
values of $p$. These are the celebrated type II D-branes~\cite{polchinski}, 
which carry charges of Ramond--Ramond gauge fields. The
dimension $p$ is even in the
IIA case and odd in the IIB case. Explicit supersymmetric 
world-volume actions with local
kappa symmetry were constructed last autumn for these D-branes
by a number of groups~\cite{cederwall1,aganagic2,bergshoeff1}.

The third case, which is the one I will focus in the remainder of this talk,
is the M5-brane. In this case the $9+5-11=3$ extra bosonic degrees of freedom
are provided by a two-form potential $B_{\mu\nu}$, whose field strength
is self dual in the linearized approximation. To understand the counting,
note that massless particles in 6d are classified by the Spin$(4) = SU(2) \times SU(2)$
little group. The states in question belong to the $(3,1)$ representation of this
group. Since parity interchanges the two $SU(2)$'s, this is a chiral boson.

The M5-brane has a simple relation to two of the $p$-branes of Type IIA
superstring theory. Recall that at strong coupling type IIA string theory
on $R^{10}$ should be reinterpreted as M theory on 
$R^{10} \times S^1$~\cite{townsend2,witten1},
where the radius of the circular 11th dimension is proportional
to the 2/3 power of the type IIA string coupling constant (in the string frame).
There are two possibilities for the fate of the M5-brane. Either one of its 
dimensions is wrapped on the circle or none of them are. In the former case,
one obtains the D4-brane and in the latter the NS solitonic fivebrane.
To understand how the D4-brane arises, one first carries out a ``double dimensional
reduction'' in which one of the world-volume coordinates is identified with the
circular spatial dimension and then the zero modes in this direction are extracted.
Next a duality transformation in the 5d world volume is required to
replace the two-form potential by the $U(1)$ gauge field that is characteristic
of the D-brane. These relationships were crucial in Witten's recent analysis
of a IIA brane configuration that corresponds to an $N=2$ $D=4$ gauge theory~\cite{witten2}.
The 10d picture involved a number of interconnected D4-branes and NS5-branes,
but the configuration could be reinterpreted as a single M5-brane in 11d.
This led to a simple geometrical interpretation of the Seiberg--Witten
Riemann surface as two of the dimensions of the M5-brane! 
(See~\cite{lerche,klemm} for reviews of SW theory and related string theoretic
approaches to understanding them geometrically.) One implication of this
result is that the M5-brane action, which we will describe, encodes a great
deal of quantum information even though all of our manipulations appear to be classical.

\medskip

\section{The Bosonic Part of the M5-Brane Action}

\subsection{Symmetries}

In a covariant formulation all global symmetries, broken and unbroken, are
exhibited. In addition, there are various local symmetries of the world-volume
theory (discussed below). When these local symmetries are used to fix a
physical gauge, the broken global symmetries become non-linearly realized
with associated Goldstone particles. Thus, the covariant M5-brane action has manifest
global 11d super-Poincar\'e symmetry realized by the superspace transformations
$\delta\theta =\epsilon$, $\delta X^M = \bar\epsilon \Gamma^M\theta + a ^M$.
However, the unbroken global symmetry of the gauge-fixed world-volume
theory is just (2,0) super-Poincar\'e symmetry in 6d. The two chiral
spinors and five scalars in the (2,0) tensor multiplet are the Goldstone 
particles associated to the broken global symmetries. 

There are three kinds of local symmetries. Two of them are essentially the
same for all super $p$-branes. They are world-volume diffeomorphism symmetry and local
kappa symmetry. The diffeomorphism symmetry implies that
the components of $X^M$ along the $p$-brane directions are
unphysical and can be gauged away, e.g. by choosing a static gauge.
Local kappa symmetry has the effect of eliminating half of the
fermionic degrees of freedom, which is essential to get the counting
required by supersymmetry. Since this paper only describes the bosonic sector of the
theory, nothing more will be said about this symmetry, except to not that
the complete kappa-symmetric action has been constructed 
recently~\cite{bandos1,aganagic3,howe1}.
The third class of local symmetries are those
that are required to properly describe a self-dual tensor. I am not referring to the
standard gauge transformation $\delta B = d \Lambda$, which is rather trivial,
but rather to the new PST gauge symmetries~\cite{pasti1,pasti2}, 
described below, which are required
to decouple the anti-chiral components of $B$.

\medskip

\subsection{Topological Issues}

In the analysis presented below I will only analyze infinitesimal diffeomorphisms
of the 6d world-volume theory. However, when the 6d manifold is topologically
nontrivial one should also analyze ``large'' diffeomorphisms, which are
not continuously connected to the identity. In other words, one should check
modular invariance. 

To explain the issues let us briefly consider an example. Suppose the 11d spacetime
is topologically of the form $R^7 \times K3$ and that the 6d world volume is
$\Sigma \times K3$, where $\Sigma$ is a Riemann surface and the world-volume
$K3$ is wrapped on the space-time $K3$. In this case one can carry out a 
double dimensional reduction obtaining an effective string action in a
7d spacetime. In fact, it was noted by Witten that M theory compactified on $K3$
is dual to the heterotic string theory compactified on $T^3$~\cite{witten1}. Furthermore,
Harvey and Strominger noted that the double dimensional reduction gives
precisely the heterotic string in 7d~\cite{harvey1}. This has been analyzed in greater detail,
using the explicit M5-brane action, by Cherkis and me~\cite{cherkis}.

Because $K3$ has 19 anti-self-dual 2-forms ($b_2^- = 19$) and three
self-dual ones  ($b_2^+ = 3$), the zero modes of the $B$ field give rise
to 19 left-moving and 3 right-moving chiral bosons on $\Sigma$. They are
compact and their momentum lattice is the even self-dual Narain lattice
$\Gamma_{19,3}$. This describes the heterotic string in 7d, which is well-known
to have modular invariance, precisely because the lattice is even and self-dual.
This strongly suggests, but does not completely prove, that the M5-brane on 
$\Sigma \times K3$ also has the requisite modular invariance. For a more
detailed discussion of these issues the reader is referred to a recent analysis
by Witten~\cite{witten3}.

\medskip

\subsection{The Noncovariant Formulation}

Ref. \cite{jhs} analyzed the problem of coupling a 6d self-dual tensor gauge field to a
metric field so as to achieve general coordinate invariance. 
It presented a formulation in which one direction is treated differently from
the other five. At the time that work was done,
the author knew of no straightforward way to make the general
covariance manifest. However, shortly thereafter a paper appeared~\cite{pasti1} that 
presents equivalent results using a manifestly covariant formulation~\cite{pasti2},
which we refer to as the PST formulation. In the following both approaches and their
relationship are described. These results have been generalized to
supersymmetric actions with local kappa symmetry~\cite{bandos1,aganagic3,howe1}, 
but here we will only consider the bosonic theories.

Let us denote the 6d world-volume coordinates by 
$\sigma^{\hat\mu} = (\sigma^\mu, \sigma^5)$,
where $\mu = 0,1,2,3,4$. The $\sigma^5$ direction is singled out as the one that
will be treated differently from the other five.\footnote{This is a
space-like direction, but one could also choose a time-like  one.}  The 6d metric
$G_{\hat\mu\hat\nu}$ contains 5d pieces $G_{\mu\nu}, G_{\mu 5}$, and $G_{55}$.
All formulas will be written with manifest 5d general coordinate invariance.
As in refs.~\cite{perry,jhs}, we represent the self-dual tensor gauge field by a
$5\times 5$ antisymmetric tensor $B_{\mu\nu}$, and its 5d curl by
$H_{\mu\nu\rho} = 3 \partial_{[\mu} B_{\nu\rho]}$. A useful quantity is the dual 
\begin{equation}
\tilde{H}^{\mu\nu} = {1\over 6} \epsilon^{\mu\nu\rho\lambda\sigma}
H_{\rho\lambda\sigma}.
\end{equation}

It was shown in ref.~\cite{jhs} that a class of generally covariant
bosonic theories can be represented in the form
$L = L_1 + L_2 + L_3$, where
\begin{eqnarray}
L_1 &=& -{1\over 2}\sqrt{-G} f(z_1,z_2), \nonumber \\
L_2 &=& -{1\over 4} \tilde{H}^{\mu\nu} \partial_5 B_{\mu\nu}, \\
L_3 &=&  {1\over 8}
\epsilon_{\mu\nu\rho\lambda\sigma} {G^{5\rho}\over G^{55}} \tilde{H}^{\mu\nu}
\tilde{H}^{\lambda\sigma}.\nonumber 
\end{eqnarray}
The notation is as follows:  $G$ is the 6d determinant $(G =
{\rm det}\, G_{\hat\mu\hat\nu})$ and
$G_5$ is the 5d determinant $(G_5 =
{\rm det}\, G_{\mu\nu})$, while $G^{55}$ and $G^{5\rho}$ are components of the inverse
6d metric $G^{\hat\mu \hat\nu}$.  The $\epsilon$ symbols are purely numerical with $\epsilon^{01234} = 1$ and
$\epsilon^{\mu\nu\rho\lambda\sigma} = - \epsilon_{\mu\nu\rho\lambda\sigma}$.  A
useful relation is $G_5 = G G^{55}$.
The $z$ variables are defined to be
\begin{eqnarray}
z_1 &=& {{\rm tr} (G\tilde{H} G\tilde{H})\over 2( -G_5)}\nonumber \\
z_2 &=& {{\rm tr} (G\tilde{H} G\tilde{H} G\tilde{H} G\tilde{H})\over 4 (-G_5)^2}.
\label{zdefs}
\end{eqnarray}
The trace only involves 5d indices:
\begin{equation}
{\rm tr} (G\tilde{H} G\tilde{H}) = G_{\mu\nu} \tilde{H}^{\nu\rho} G_{\rho\lambda}
\tilde{H}^{\lambda\mu}.
\end{equation}
The quantities $z_1$ and $z_2$
are scalars under 5d general coordinate
transformations.  

Infinitesimal parameters of general coordinate transformations are denoted
$\xi^{\hat\mu} = (\xi^\mu, \xi)$.  Since 5d general coordinate invariance is
manifest, we focus on the $\xi$ transformations only.  The metric transforms in
the standard way
\begin{equation}
\delta_\xi G_{\hat\mu \hat\nu} = \xi \partial_5 G_{\hat\mu \hat\nu} +
\partial_{\hat\mu} \xi G_{5\hat\nu} + \partial_{\hat\nu} \xi G_{\hat\mu 5}.
\label{Gvar}
\end{equation}
The variation of $B_{\mu\nu}$ is given by a more complicated rule, whose origin is
explained in ref.~\cite{jhs}:
\begin{equation}
\delta_\xi B_{\mu\nu} = \xi K_{\mu\nu}, \label{Bvar}
\end{equation}
where
\begin{equation}
K_{\mu\nu} = 2{\partial (L_1 + L_3) \over\partial \tilde{H}^{\mu\nu}} =
K_{\mu\nu}^{(1)} f_1+ K_{\mu\nu}^{(2)} f_2+ K_{\mu\nu}^{(\epsilon)}
\label{Kform1}
\end{equation}
with
\begin{eqnarray}
K_{\mu\nu}^{(1)} &=& {\sqrt{-G} \over (-G_5)}{(G\tilde{H} G)_{\mu\nu}}
\nonumber \\
K_{\mu\nu}^{(2)} &=& {\sqrt{-G} \over (-G_5)^2}{(G\tilde{H} G\tilde{H} G\tilde{H}
G)_{\mu\nu}}  \label{Kform2}\\
K_{\mu\nu}^{(\epsilon)} &=& \epsilon_{\mu\nu\rho\lambda\sigma}
{G^{5\rho}\over 2 G^{55}} \tilde{H}^{\lambda\sigma}, \nonumber
\end{eqnarray}
and we have defined
\begin{equation}
f_i = {\partial f\over\partial z_i} , \quad i = 1,2.
\end{equation}

Assembling the results given above, ref.~\cite{jhs} showed that
the required general coordinate transformation symmetry is
achieved if, and only if, the function $f$ satisfies the nonlinear partial
differential equation~\cite{gibbons}
\begin{equation}
f_1^2 + z_1 f_1 f_2 + \big({1\over 2} z_1^2 - z_2\big) f_2^2 = 1.
\end{equation}
As discussed in~\cite{perry},
this equation has many solutions, but the one of relevance to the
M theory five-brane is 
\begin{equation}
f = 2 \sqrt{1 + z_1 + {1\over 2} z_1^2 - z_2}.
\end{equation}
For this choice $L_1$
can reexpressed in the Born--Infeld form
\begin{equation}
L _1 = - \sqrt{- {\rm det} \Big(G_{\hat\mu \hat\nu} + i G_{\hat\mu\rho} G_{\hat\nu
\lambda} \tilde{H}^{\rho\lambda} / \sqrt{-G_5}\Big)} . \label{bosonicL1}
\end{equation}
This expression is real, despite the factor of $i$, because it is an even function of
$\tilde H$.

\medskip

\subsection{The PST Formulation}

In ref.~\cite{pasti1} (using techniques developed
in ref.~\cite{pasti2}) equivalent results are
described in a manifestly covariant way.  To do this, the field $B_{\mu\nu}$ is
extended to $B_{\hat\mu \hat\nu}$ with field strength $H_{\hat\mu \hat\nu
\hat\rho}$.  In addition, an auxiliary scalar field $a$ is
introduced.  The PST formulation has new gauge symmetries (described below)
that allow one to choose the gauge $B_{\mu 5} = 0,$  $a = \sigma^5$
(and hence $\partial_{\hat\mu}a =
\delta_{\hat\mu}^5$).  In this gauge, the covariant PST formulas reduce to the
ones given above.

Equation (\ref{bosonicL1}) 
expresses $L_1$ in terms of the determinant of the $6 \times 6$ matrix
\begin{equation}
M_{\hat\mu\hat\nu} = G_{\hat\mu\hat\nu} + i {G_{\hat\mu \rho} G_{\hat\nu
\lambda}\over \sqrt{- GG^{55}}} \tilde{H}^{\rho\lambda}.
\end{equation}
In the PST approach this is extended to the manifestly covariant form
\begin{equation}
M_{\hat\mu\hat\nu}^{\rm cov.} = G_{\hat\mu\hat\nu} + i {G_{\hat\mu\hat\rho}
G_{\hat\nu \hat\lambda}\over\sqrt{-G (\partial a)^2}}
\tilde{H}_{\rm cov.}^{\hat\rho \hat\lambda}. \label{Mcov}
\end{equation}
The quantity
\begin{equation}
(\partial a)^2 = G^{\hat\mu\hat\nu} \partial_{\hat\mu} a \partial_{\hat\nu} a
\end{equation}
reduces to $G^{55}$ upon setting $\partial_{\hat\mu}a  = \delta_{\hat\mu}^5$,
and
\begin{equation}
\tilde{H}_{\rm cov.}^{\hat\rho\hat\lambda} \equiv {1\over 6} \epsilon^{\hat\rho
\hat\lambda \hat\mu \hat\nu \hat\sigma \hat\tau} H_{\hat\mu \hat\nu \hat\sigma}
\partial_{\hat\tau} a
\end{equation}
reduces to $\tilde{H}^{\rho\lambda}$.  Thus $M_{\hat\mu \hat\nu}^{\rm cov.}$
replaces $M_{\hat\mu\hat\nu}$ in $L_1$.  Furthermore, the expression
\begin{equation}
L' = - { 1\over 4(\partial a)^2} \tilde{H}_{\rm cov.}^{\hat\mu \hat\nu}
H_{\hat\mu\hat\nu\hat\rho} G^{\hat\rho\hat\lambda} \partial_{\hat\lambda} a,
\end{equation}
which transforms under general coordinate transformations as a scalar density,
reduces to $L_2 + L_3$ upon gauge fixing. It is interesting that $L_2$ and $L_3$ are
unified in this formulation.

Let us now describe the new gauge symmetries of ref.~\cite{pasti1}.  Since degrees of
freedom $a$ and $B_{\mu 5}$ have been added, corresponding gauge symmetries are
required.  One of them is
\begin{equation}
\delta B_{\hat\mu \hat\nu} = 2 \phi_{[\hat\mu} \partial_{\hat\nu]} a,
\end{equation}
where $\phi_{\hat\mu}$ are infinitesimal parameters, and the other fields do not
vary.  In terms of differential forms, this implies $\delta H = d\phi\wedge
da$.  $\tilde{H}_{\rm cov.}^{\hat\rho \hat\lambda}$ is invariant under this transformation,
since it corresponds to the dual of $H\wedge da$, but $da\wedge da = 0$.
Thus the covariant version of $L_1$ is invariant under this transformation.
The variation of $L'$, on the other hand, is a total derivative.

The second local symmetry involves an infinitesimal 
scalar parameter $\varphi$.  The transformation
rules are $\delta G_{\hat\mu\hat\nu} = 0, \delta a = \varphi$, and
\begin{equation}
\delta B_{\hat\mu\hat\nu} = {1\over (\partial a)^2} \varphi
H_{\hat\mu\hat\nu\hat\rho} G^{\hat\rho\hat\lambda} \partial_{\hat\lambda} a +
\varphi V_{\hat\mu\hat\nu},
\end{equation}
where the quantity $V_{\hat\mu\hat\nu}$ is to be determined.   
Rather than derive it from
scratch, let's see what is required to agree with the previous formulas after
gauge fixing.  In other words, we fix the gauge $\partial_{\hat\mu} a =
\delta_{\hat\mu}^5$ and $B_{\mu 5} = 0$, and figure out what the resulting
$\xi$ transformations are.  We need
\begin{equation}
\delta a = \varphi + \xi \partial_5 a = \varphi + \xi = 0,
\end{equation}
which tells us that $\varphi = - \xi$.  Then
\begin{eqnarray}
\delta_{\xi} B_{\mu\nu} &=& {1\over (\partial a)^2} \varphi H_{\mu\nu\hat\rho}
G^{\hat\rho\hat\lambda} \partial_{\hat\lambda} a + \varphi V_{\mu\nu} + \xi
H_{5\mu\nu}\nonumber \\
&=& - \xi \left({G^{\rho 5}\over G^{55}} H_{\mu\nu\rho} + V_{\mu\nu}\right) =
\xi (K_{\mu\nu}^{(\epsilon)} - V_{\mu\nu}).
\end{eqnarray}
Thus, comparing with eqs.~(\ref{Bvar}) and (\ref{Kform1}), we need the covariant definition
\begin{equation}
V_{\hat\mu\hat\nu} = - 2 {\partial L_1\over \partial
\tilde{H}_{\rm cov.}^{\hat\mu\hat\nu}}
\end{equation}
to achieve agreement with our previous results.

The presence of the factor $(\partial a)^2$ in various denominators is potentially
problematical unless the six manifold on which the action is defined satisfies
an appropriate topological condition. The precise determination of what that condition
is, and how it relates to the condition of modular invariance discussed earlier,
deserve further study.

\section{Discussion}

The explicit formula for the M5-brane action should be useful for a number of purposes.
For one thing, Witten's derivation~\cite{witten2} of the Seiberg--Witten quantum effective
action only required knowledge of the quadratic (free) approximation. 
The complete nonlinear formula should allow one to
extend the analysis to derive some of the higher dimension terms in the 4d
quantum effective action. Exactly which ones 
could be derived is not entirely clear, but I suppose
that the requisite mathematical control would be best for those that are encoded in
holomorphic functions.

Evidence has been mounting that there are several classes of 
interacting non-gravitational 6d quantum
theories. Recently, Seiberg has argued that
starting from a set of $k$ parallel M5-branes one can define a class of these theories
by means of a suitable limit in which they decouple from the bulk degrees of freedom
while still remaining interacting~\cite{seiberg}. 
Such a theory is certainly not a conventional
quantum field theory, and it is not yet entirely clear how it should best be
formulated. In any case, if $k>1$ there are string-like excitations due to
supermembranes attaching between a pair of the fivebranes. 
If $k=1$, and if one of the transverse
dimensions is compact, so that there is IIA interpretation, then there are
also strings due to a supermembrane that starts and ends on the fivebrane
and wraps around the circular transverse dimension.  These considerations
have led to suggestions that these might be viewed as ``tensionless'' or
``noncritical'' string theories. In any case, whatever the proper fundamental
formulation will turn out to be, the fomulas we have obtained should be
viewed as the low-energy effective action for the $k=1$ case, just as
11d supergravity is the low-energy effective description of M theory. In fact,
if these 6d theories can be formulated as matrix models, the analogy would
be a good one.

\section*{Acknowledgment}

I am grateful to M. Aganagic, S. Cherkis, 
J. Park, M. Perry, and C. Popescu for collaborating on portions of this work.
This work is 
supported in part by the U.S. Dept. of Energy under Grant No.
DE-FG03-92-ER40701.

\end{document}